# Approaching the type-II Dirac point and concomitant superconductivity in Pt-doping stabilized metastable 1T-phase IrTe$_2$

Fucong Fei[1†], Xiangyan Bo[1†], Pengdong Wang[2], Jianghua Ying[3], Bo Chen[1], Qianqian Liu[1], Yong Zhang[1], Zhe Sun[2], Fanming Qu[3], Yi Zhang[1], Jian Li[4], Fengqi Song[1*], Xiangang Wan[1*], Baigeng Wang[1*], Guanghou Wang[1]

[1] *National Laboratory of Solid State Microstructures, Collaborative Innovation Center of Advanced Microstructures, and College of Physics, Nanjing University, Nanjing 210093, China*

[2] *National Synchrotron Radiation Laboratory, University of Science and Technology of China, Hefei 230029, China*

[3] *Beijing National Laboratory for Condensed Matter Physics, Institute of Physics, Chinese Academy of Sciences, Beijing 100190, China*

[4] *Westlake Institute for Advanced Study, Hangzhou 310012, China*

[†] These authors contributed equally to this work.

[*] e-mail: songfengqi@nju.edu.cn; xgwan@nju.edu.cn; bgwang@nju.edu.cn

**Abstract**

Topological semimetal is a topic of general interest in material science. Recently, a new kind of topological semimetal called type-II Dirac semimetal with tilted Dirac cones is discovered in $PtSe_2$ family. However, the further investigation is hindered due to the huge energy difference from Dirac points to Fermi level and the irrelevant conducting pockets at Fermi surface. Here we characterize the optimized type-II Dirac dispersions in a metastable 1T phase of $IrTe_2$. Our strategy of Pt doping protects the metastable 1T phase in low temperature and tunes the Fermi level to the Dirac point. As demonstrated by angle-resolved photoemission spectra and first principle calculations, the Fermi surface of $Ir_{1-x}Pt_xTe_2$ is formed by only a single band with type-II Dirac cone which is tilted strongly along $k_z$ momentum direction. Interesting superconductivity is observed in samples for Dirac point close to Fermi level and even survives when Fermi level aligns with the Dirac point as finite density of states created by the tilted cone dispersion. This advantage offers opportunities for possible topological superconductivity and versatile Majorana devices in type-II Dirac semimetals.

Topological semimetals (TSM) have generated wide spread interest in materials science. Multiple series of topological nontrivial TSM are widely researched including Dirac semimetals (DSM)[1-4], Weyl semimetals (WSM)[5-7], nodal-line semimetals[8,9] as well as Lorentz invariance breaking type-II WSMs[10-12]. Recently, the concept of broken Lorentz invariance is also introduced to DSM resulting a new emerging system named type-II DSM, where the Dirac cones are greatly tilted in a momentum direction[13-20], leading to modulated effective mass and versatile device opportunities[21-23]. In case of coexisting superconductivity, topological superconductivity (TSC)[24-28] may be expected with the potential future of Majorana fermion[24-32] and topological quantum computation[33,34], where broader momentum dispersion hosts more freedoms. The real materials of type-II DSM are discovered in 1T-phase $PtSe_2$ family, i. e. $PtSe_2$, $PtTe_2$ and $PdTe_2$, recently[13,15-18]. All these materials are group-10 transition-metal dichalcogenides (TMDC) with similar structures with $P\bar{3}m1$ space group and offers very similar electronic structures of type-II Dirac dispersions which is protected by $C_{3v}$ rotation symmetry[13,14].

In spite of wide researches, all these materials are plagued by an issue that the Dirac points are far below the Fermi level and there are several trivial bands crossing the Fermi surface[13-18], which hinders the materials from transport studies and future application. Thus it is crucial to discover a type-II DSM system with Dirac points closer to Fermi level as well as kill the trivial pockets. As the Dirac dispersion is protected by crystal symmetry, substituting group-9 transition metals, e. g. iridium, for Pt/Pd and forming $IrTe_2$ with 1T-phase can pull the Dirac point closer to Fermi surface with keeping the Dirac dispersion. However, as what commonly happens in TMDCs, 1T-IrTe$_2$ is metastable and a charge density wave (CDW) transition to monoclinic structure occurs in low temperature. Thus, Dirac dispersion in IrTe$_2$ no longer survives[35-40].

Here we report our strategy of platinum doping in metastable 1T phase of IrTe$_2$, which not only stabilize the metastable 1T-phase by suppress the CDW transition but also tunes the Fermi level to the Dirac point by introducing carriers. As demonstrated by angle-resolved photoemission spectra (ARPES) and first principle calculations, the type-II Dirac point approaches to Fermi level by adjusting doping strength as expected and a single band with Dirac cone is formed at Fermi surface. Interesting superconductivity is observed in samples for Dirac point close to Fermi level. This offers the opportunities for possible TSC and mysterious Majorana fermions.

Our strategy is first demonstrated by the first-principle calculations. We calculated the band structure of IrTe$_2$ (Fig. 1a-b). The crystal structure and Brillouin zone (BZ) are shown in Fig. 1c and d respectively. There are two bands cross the Fermi level, forming a band crossing feature near the Fermi level along the Γ-A direction (marked by the red arrows). As these two bands belong to different representations (G$_4$ and G$_{5,6}$ respectively), the spin-orbital coupling (SOC) cannot open a gap at cross point and instead a gapless Dirac cone with linear dispersion is formed. This is determined by the C$_{3v}$ symmetry in the system with inversion symmetry and time-reversal symmetry[13,14,41]. It is obvious from Fig. 1a and b that the Dirac cone is untilted in k$_x$-k$_y$ plane (T-D-S direction) but tilted strongly along k$_z$ direction (A-Γ-A direction), which is the characteristic of the type-II Dirac fermions. According to our calculations, the Dirac point is 200meV higher than Fermi level, much closer than PtSe$_2$ family[14]. If some of the Ir atoms can be substitute by Pt, as discussed above, to make an Ir$_{1-x}$Pt$_x$Te$_2$ sample, The Dirac point will drop down and align with Fermi level at a certain doping level. In addition to the tunable Fermi level, an ideal type-II Dirac semimetal with single Dirac transport will be held in Ir$_{1-x}$Pt$_x$Te$_2$ materials. We delightedly found that there are only two bands which forming the Dirac cone cross the Fermi level in IrTe$_2$ and other irrelevant bands are all far away from Fermi surface (Fig. 1a-b). This means for a wide range of doping level x, the Fermi surface in Ir$_{1-x}$Pt$_x$Te$_2$ is formed by a single band with Dirac cone. As a character of type-II Dirac dispersion, a pair of hole-like and electron-like pockets will be formed by the tilted Dirac cone. Fig. 1e and f are the constant energy surface of Fermi level and 0.2eV higher than Fermi level (i.e., the energy of Dirac point) respectively and the pair of hole-like (red color) and electron-like (blue) pockets can be clearly seen. In Fig. 1f, the hole-like pocket is located at the center of BZ. The shape of electron-like pocket is complex and divided into two parts by the boundary of BZ. The outer part has the lantern shape and traverses the BZ along k$_z$ direction, while the inner part is ellipsoid shaped and touches with hole-like pocket at Dirac points (indicate by the black arrows). No other irrelevant pockets can be found both in Fig. e and f, the single band of Dirac cone crossing the Fermi level is thus further confirmed.

Single crystal Ir$_x$Pt$_{1-x}$Te$_2$ with different x was grown by self-flux method. Millimeter-sized crystals with hexagonal shape were obtained (inset in Fig. 2a), which could be exfoliated by a knife easily. Fig. 2a shows the powder x-ray diffraction data of one typical $Ir_{1-x}Pt_xTe_2$ sample. As there's no standard diffraction card of these doped samples to consult to, we use the crystal model with $P\bar{3}m1$ space group to fit the experimental data. The fitting curve (black curve) is consistent with the experimental data (red dots), indicating the pure 1T phase of $Ir_{1-x}Pt_xTe_2$ samples in room-temperature. Fig. 2b shows the energy dispersive spectra (EDS) of one typical sample and the inset shows part of the EDS which correspond to the La peaks of Ir and Pt from samples with different x. The peaks height is normalized by the peaks from Te around 4 keV. The ratio of two elements changes clearly and indicates the successful control of doping in $Ir_{1-x}Pt_xTe_2$.

As discussed above, a CDW phase transition may occur in this system and kills the 1T-phase as well as the Dirac dispersions when the temperature decreases, it is necessary to check whether the 1T phase still survives in our samples under low temperature. Fig. 2c displays the resistance versus temperature of four $Ir_{1-x}Pt_xTe_2$ samples with different x from 0.1-0.4. The cooling curve and the heating curve of each sample are perfectly coincident and no sudden jump occurred, indicating no crystal phase transition in these samples. Combining with the XRD results, the stability of 1T-phase is confirmed in our samples. In addition, one may notice that the sample of x=0.1 (black curve) shows superconducting behavior with $T_c$=1.7K, which will be further discussed below.

After the 1T phase was confirmed in $Ir_{1-x}Pt_xTe_2$, we performed ARPES measurements to detect the band dispersions in our samples and compare with the

DFT calculation. Fig. 3 a-c are the band dispersions along K-Γ-M direction of $Ir_{1-x}Pt_xTe_2$ with different x. The black dash lines are theoretical band calculation of $IrTe_2$. The Fermi energy of calculation in each panel is shifted and the calculation data is consistent with the experimental data. The offset of Fermi level in different samples is obvious. For the sample of x=0.2, the Dirac point is a little bit higher than Fermi level (~0.05eV). For the sample of x=0.3, the Dirac point is aligned with Fermi level. For the sample of x=0.4, as more Pt atoms inject more electrons into the solid, the Dirac point is about 0.2eV lower than Fermi level and the upper part of the Dirac cone can be detected by ARPES. The artificial tuning of Fermi level near Dirac cone in $Ir_{1-x}Pt_xTe_2$ by changing the dopant strength is thus confirmed. We also measured the constant energy contours in our sample. The upper panel in Fig. 3d shows the constant energy mappings of x=0.2 sample measured at Fermi energy to 1eV lower than Fermi energy. The corresponding theoretical calculations of $IrTe_2$ after Fermi energy shift are displayed in the lower panel and well agree with the experiment. One may also notice that in Fig. 3a there is a cross-like band marked by the yellow dot lines (also in Fig 3b and c but obscure). We believe this is a surface Dirac cone state which is consistent with previous reports[13,42].

To reveal the bulk properties of the Dirac cone in $Ir_{1-x}Pt_xTe_2$, we use various photon energies to collect the $k_z$ dispersions of our samples. Fig. 4b-g show the dispersions of sample with x=0.2 measured at photon energy from 18.5eV to 23.5eV and the dash lines are theoretical band calculation of $IrTe_2$ (after Fermi energy shifting) at corresponding $k_z$ momentum. The $k_z$ values are calculated with inner potential of

13.5eV. The evolution of the Dirac cone dispersion can be clearly seen which no doubt reveals the bulk properties of the Dirac cone. While the surface Dirac cone dispersions (marked by the yellow dot lines) keep constant except the changing of signal intensity. In addition, we can also reveal the type-II properties of the bulk Dirac cone by the ARPES data under different photon energies. Fig. 4a is a sketch of type-I and type-II Dirac cones. Let's consider the valence band forming the lower part of the Dirac cone (blue color). The top positions of the valence band (Γ point in our case) under different momentum ($k_z$ in our case) are marked by the red dots. One can clearly see for a type-I Dirac cone, the top position is highest at Dirac point and reduces when departing from Dirac point. However, for a type-II Dirac cone, the top position goes monotonically when momentum sweeps over the tilted cone region. In our ARPES data, the valence band forming Dirac cone in each panel is marked by the red dash line. The position of the valence band at Γ point under photon energy of 18.5eV and 19.5eV is somewhere upper than Fermi level and becomes lower when increasing the photon energy, i. e. increasing $k_z$, which fit the case of type-II Dirac cone. Therefore, the type-II property of Dirac cone in $Ir_{1-x}Pt_xTe_2$ is verified by our experiment.

After demonstrating the ideal type-II Dirac dispersion with single band with Dirac cone crossing the Fermi level as well as the artificially tunable Dirac point position by doping in our $Ir_{1-x}Pt_xTe_2$ samples, processing measurement of transport properties, such as superconductivity, in these materials becomes particularly important. Hence we carry out further investigation of the low-temperature transport properties in our $Ir_{1-x}Pt_xTe_2$ samples. There are several previous reports[36,38,40] on the superconductivity in $Ir_{1-x}Pt_xTe_2$, however, the Pt doping level is far lower than our samples. It is generally considered that superconductivity is no longer existed when x>0.1. We measure the transport of our samples under ultra-low temperature by a dilution refrigerator and surprisingly find that superconductivity actually survives in $Ir_{1-x}Pt_xTe_2$ with wide range of x as shown in Fig. 5a. For sample with x = 0.1, the superconducting transition temperature ($T_c$) is around 1.7 K, which is consistent with the previous report[36]. For sample with x = 0.2, the $T_c$ decreases apparently to 0.65 K. When the doping level x keeps increasing to 0.3, the superconductivity is still survived but $T_c$ drops to only 0.15 K. For samples with x=0.4, the superconductivity is finally killed in the temperature range we can get (>47 mK). Temperature dependence of the upper critical fields $B_{c2}$ of the samples with different x are displayed in Fig. 5b by solid dots and the smooth curves are the best fit curves by using equation: $B_{C2}(T) = B_{C2}^*(1-T/T_c)^{1+\alpha}$. We summary all the obtained results in the superconductivity phase diagram displayed in Fig. 5c. When x < 0.04, a phase transition to monoclinic structure occurs in $Ir_{1-x}Pt_xTe_2$ under low temperature[36]. Increasing Pt doping strength, metastable 1T-phase is stabilized along with superconductivity emerges in low temperature. When further increasing x, $T_c$ decreases and superconductivity disappears for x > 0.3.

Taking the relationship between Dirac point and Fermi energy for corresponding sample obtained by ARPES into consideration (displayed by the sketch in Fig. 5 c), we are surprised to find that the Dirac points for those samples displaying superconductivity are close to Fermi level. It is worth noting that superconductivity even survives in the sample for Fermi level aligning with the Dirac point (x=0.3). This may explain by the unique tilted cone dispersions in type-II semimetals. It is well known that in traditional type-I DSM there is little density of state (DOS) near Dirac points. This may eliminate the possible superconductivity near the Dirac point where the topological supercurrent is expected to the best. Contrary to the vanishing of DOS at Dirac point in type-I DSMs, finite DOS at Dirac point is created by the pair of electron and hole pockets formed by the tilted cone dispersion[10] in type-II DSM and superconductivity can be thus promoted at Dirac point. We believe this may explain the survived superconductivity in samples for Fermi level aligning with the Dirac point. This superconductivity promoting advantage may offer the opportunities for possible topological superconductivity and searching for Majorana fermions in the future.

## Methods

**Crystal growth.** Single crystal $Ir_xPt_{1-x}Te_2$ with different x was grown by self-flux method. A certain amount of iridium powder (99.9%, aladdin), platinum sponge (99.98%, Alfa Aesar) and tellurium shot (99.99%, aladdin) with an atomic ratio 0.18(1-x):0.18x:0.82 were placed in an alumina crucible. The crucible was vacuum sealed in a thick-walled ampoule. The ampoule was heated up and kept at 950℃ for 10 hours, then at 1160℃ for 24 hours. The melt was then slowly cooled down to 880 ℃ in a week. Then the Te flux was separated from single crystals by centrifugation.

**Angle-resolved photoemission spectroscopy.** ARPES measurements were performed at BL13U of Hefei National Synchrotron Radiation Laboratory. The samples were in-situ cleaved and measured in the chamber with pressure below than $10^{-10}$ Torr. Photon energies from 15 eV-30 eV were used for measurement.

**Theoretical calculation.** The electronic structure calculations were carried out by using the full potential linearized augmented planewave method as implemented in the WIEN2K package[43]. The General Gradient Approximate for the correlation potential has been used here. Using the second-order variational procedure, we include the spin-orbital coupling interaction. The lattice constants we used here are a = b = 3.93 Å and c = 5.39 Å taking a space group of $P\bar{3}m1$. The muffin-tin radii for Ir and O are set to 1.32 Å and 1.27 Å, respectively. The basic functions are expanded to $R_{mt}K_{max} = 7$ (where $R_{mt}$ is the smallest of the muffin-tin sphere radii and $K_{max}$ is the largest reciprocal lattice vector used in the plane-wave expansion). A 14 × 14 × 9 k-point mesh is used for the Brillouin zone integral. The self-consistent calculations are considered to be converged when the difference in the total energy of the crystal does not exceed 0.1 mRy.

**References**

1  Liu, Z. K. *et al.* A stable three-dimensional topological Dirac semimetal Cd3As2. *Nature Mater.* **13**, 677-681, (2014).
2  Liang, T. *et al.* Ultrahigh mobility and giant magnetoresistance in the Dirac semimetal Cd3As2. *Nat Mater* **14**, 280-284, (2015).
3  Xu, S.-Y. *et al.* Observation of Fermi arc surface states in a topological metal. *Science* **347**, 294-298, (2015).

4   Liu, Z. K. *et al.* Discovery of a Three-Dimensional Topological Dirac Semimetal, Na3Bi. *Science* **343**, 864-867, (2014).

5   Wan, X., Turner, A. M., Vishwanath, A. & Savrasov, S. Y. Topological semimetal and Fermi-arc surface states in the electronic structure of pyrochlore iridates. *Phys. Rev. B* **83**, 205101, (2011).

6   Yang, L. X. *et al.* Weyl semimetal phase in the non-centrosymmetric compound TaAs. *Nature Phys.* **11**, 728–732, (2015).

7   Xu, S.-Y. *et al.* Discovery of a Weyl fermion semimetal and topological Fermi arcs. *Science* **349**, 613-617, (2015).

8   Hu, J. *et al.* Evidence of Topological Nodal-Line Fermions in ZrSiSe and ZrSiTe. *Phys. Rev. Lett.* **117**, 016602, (2016).

9   Pezzini, S. *et al.* Unconventional mass enhancement around the Dirac nodal loop in ZrSiS. *Nature Phys.*, (2017).

10  Soluyanov, A. A. *et al.* Type-II Weyl semimetals. *Nature* **527**, 495-498, (2015).

11  Wang, Z. *et al.* MoTe2: A Type-II Weyl Topological Metal. *Phys. Rev. Lett.* **117**, 056805, (2016).

12  Xu, S.-Y. *et al.* Discovery of Lorentz-violating type II Weyl fermions in LaAlGe. *Sci. Adv.* **3**, e1603266, (2017).

13  Bahramy, M. S. *et al.* Ubiquitous formation of bulk Dirac cones and topological surface states from a single orbital manifold in transition-metal dichalcogenides. *Nature Mater.*, Published online: 27 November 2017, doi:2010.1038/nmat5031.

14  Huang, H., Zhou, S. & Duan, W. Type-II Dirac Fermions in the PtSe2 class of transition metal dichalcogenide. *Phys. Rev. B* **94**, 121117(R), (2016).

15  Fei, F. *et al.* Nontrivial Berry phase and type-II Dirac transport in the layered material PdTe2. *Phys. Rev. B* **96**, 041201, (2017).

16  Yan, M. *et al.* Lorentz-violating type-II Dirac fermions in transition metal dichalcogenide PtTe2. *Nat. Commun.* **8**, 257, (2017).

17  Noh, H.-J. *et al.* Experimental Realization of Type-II Dirac Fermions in a PdTe2 Superconductor. *Phys. Rev. Lett.* **119**, 016401, (2017).

18  Zhang, K. *et al.* Experimental evidence of type-II Dirac fermions in PtSe2. *arXiv*, 1703.04242, (2017).

19  Chang, T. R. *et al.* Type-II Symmetry-Protected Topological Dirac Semimetals. *Phys. Rev. Lett.* **119**, 026404, (2017).

20  Guo, P.-J., Yang, H.-C., Liu, K. & Lu, Z.-Y. Type-II Dirac semimetals in the YPd2Sn class. *Phys. Rev. B* **95**, (2017).

21  O'Brien, T. E., Diez, M. & Beenakker, C. W. Magnetic Breakdown and Klein Tunneling in a Type-II Weyl Semimetal. *Phys. Rev. Lett.* **116**, 236401, (2016).

22  Xiong, J. *et al.* Evidence for the chiral anomaly in the Dirac semimetal Na3Bi *Science* **350**, 413-416, (2015).

23  Udagawa, M. & Bergholtz, E. J. Field-Selective Anomaly and Chiral Mode Reversal in Type-II Weyl Materials. *Phys. Rev. Lett.* **117**, 086401, (2016).

24  Qi, X. L., Hughes, T. L., Raghu, S. & Zhang, S. C. Time-reversal-invariant topological superconductors and superfluids in two and three dimensions.

*Phys. Rev. Lett.* **102**, 187001, (2009).

25  Fu, L. & Kane, C. L. Superconducting proximity effect and majorana fermions at the surface of a topological insulator. *Phys. Rev. Lett.* **100**, 096407, (2008).

26  He, J. J., Ng, T. K., Lee, P. A. & Law, K. T. Selective equal-spin Andreev reflections induced by Majorana fermions. *Phys. Rev. Lett.* **112**, 037001, (2014).

27  Qi, X.-L., Hughes, T. L. & Zhang, S.-C. Chiral topological superconductor from the quantum Hall state. *Phys. Rev. B* **82**, 184516, (2010).

28  Qi, X.-L. & Zhang, S.-C. Topological insulators and superconductors. *Rev. Mod. Phys.* **83**, 1057-1110, (2011).

29  Wang, M.-X. *et al.* The Coexistence of Superconductivity and Topological Order inbthe Bi2Se3 Thin Films. *Science* **336**, 52-55, (2012).

30  Du, G. *et al.* Drive the Dirac electrons into Cooper pairs in SrxBi2Se3. *Nat. Commun.* **8**, 14466, (2017).

31  Xu, S.-Y. *et al.* Momentum-space imaging of Cooper pairing in a half-Dirac-gas topological superconductor. *Nature Phys.* **10**, 943-950, (2014).

32  Shen, J. *et al.* Nematic topological superconducting phase in Nb-doped Bi2Se3. *npj Quantum Materials* **2**, 59, (2017).

33  Alicea, J., Oreg, Y., Refael, G., von Oppen, F. & Fisher, M. P. A. Non-Abelian statistics and topological quantum information processing in 1D wire networks. *Nature Phys.* **7**, 412-417, (2011).

34  Nayak, C., Simon, S. H., Stern, A., Freedman, M. & Das Sarma, S. Non-Abelian anyons and topological quantum computation. *Rev. Mod. Phys.* **80**, 1083-1159, (2008).

35  Eom, M. J. *et al.* Dimerization-induced Fermi-surface reconstruction in IrTe2. *Phys. Rev. Lett.* **113**, 266406, (2014).

36  Pyon, S., Kudo, K. & Nohara, M. Emergence of superconductivity near the structural phase boundary in Pt-doped IrTe2 single crystals. *Physica C: Superconductivity* **494**, 80-84, (2013).

37  Toriyama, T. *et al.* Switching of Conducting Planes by Partial Dimer Formation in IrTe2. *J. Phys. Soc. Jpn.* **83**, 033701, (2014).

38  Fang, A. F., Xu, G., Dong, T., Zheng, P. & Wang, N. L. Structural phase transition in IrTe2: a combined study of optical spectroscopy and band structure calculations. *Sci. Rep.* **3**, 1153, (2013).

39  Mochiku, T. *et al.* Crystal Structure of Pt-doped IrTe2 Superconductor. *Physics Procedia* **58**, 90-93, (2014).

40  Yang, J. J. *et al.* Charge-orbital density wave and superconductivity in the strong spin-orbit coupled IrTe2:Pd. *Phys. Rev. Lett.* **108**, 116402, (2012).

41  Yang, B. J. & Nagaosa, N. Classification of stable three-dimensional Dirac semimetals with nontrivial topology. *Nat. Commun.* **5**, 4898, (2014).

42  Kong, W.-D. *et al.* Surface State Bands in Superconducting (PtxIr1−x)Te2. *Chinese Phys. Lett.* **32**, 077402, (2015).

43  P., B., K., S., Madsen, G. K. H., D., K. & J., L. WIEN2K, An Augmented Plane Wave+ Local Orbitals Program for Calculating Crystal Properties

(Karlheinz Schwarz, Technische Universitat Wien, Austria, 2001).

**Acknowledgements**

We gratefully acknowledge the financial support of the National Key R&D Program of China (2017YFA0303203), National Key Projects for Basic Research of China (2013CB922103), the National Natural Science Foundation of China (91421109, 91622115, 11522432, 61176088, 11274003), the Natural Science Foundation of Jiangsu Province (BK20160659, BK20130054), the Fundamental Research Funds for the Central Universities, and the opening Project of Wuhan National High Magnetic Field Center. We also thank Prof. qianghua Wang in Nanjing University, China for stimulating discussions.

**Author contributions**

F. S. and F. F. designed the project. F. F. and B. C. took charge of crystal growth and characterization. X. W. and X. B. made the first-principles calculations. P. W., Z. S., F. F., B. C. Q. L. and Y. Z. carried out the ARPES measurement. J. Y. and F. Q. measured the transport properties. G. W., Y. Z. and J. L. participated in the discussion on this topic. F. F., F. S., X. W. and B. W. wrote a manuscript. All of the authors discussed the results and contributed to the production of the manuscript.

**Competing financial interests:** The authors declare no competing financial interests.

## Figure Captions

**Figure 1| Ideal type-II Dirac band structures in Ir$_{1-x}$Pt$_x$Te$_2$.** **a, b,** The band calculation of 1T-IrTe$_2$. **c,** The crystal structure of 1T-IrTe$_2$. **d,** The Brillouin zone of IrTe$_2$. **e, f,** The constant energy surface of Fermi level and 0.2eV higher than Fermi level (i.e., the energy of Dirac point), respectively. The pair of electron (blue) and hole (red) pockets formed by the tilted Dirac cone touch each other at Dirac points and no other irrelevant pockets can be found

**Figure 2| Crystal growth and characterization. a,** The powder XRD pattern of Ir$_{1-x}$Pt$_x$Te$_2$ with x=0.4 sample. Red dots are the raw data and the black curve is the fitting curve with crystal structure of $P\bar{3}m1$ space group. Blue curve shows the difference between raw data and fitting. Green rods display the Bragg peak positions of the fitting curve. **b,** The EDS spectrum of one typical sample. The inset shows the part of the EDS spectra which correspond to the Lα peaks of Ir and Pt from samples with different x. **c,** Resistance versus temperature of four Ir$_{1-x}$Pt$_x$Te$_2$ samples with different x. The curves are offset for clearness.

**Figure 3| Dispersions with various doping measured by ARPES. a-c,** The band dispersions along K-Γ-M of Ir$_{1-x}$Pt$_x$Te$_2$ with different x from 0.2-0.4. The black dash lines are theoretical band calculation of IrTe$_2$ (the Fermi energy is shifted for matching the ARPES measurement). **d,** Constant energy mappings of x=0.2 sample measured at Fermi energy to 1eV lower than Fermi energy (upper panel) and the

corresponding theoretical calculation of IrTe$_2$ (lower panel, the Fermi energy is shifted up by 0.13eV).

**Figure 4| Verifying type-II Dirac property in Ir$_{1-x}$Pt$_x$Te$_2$. a,** A sketch of type-I and type-II Dirac cone. **b-g,** The band dispersions of Ir$_{0.8}$Pt$_{0.2}$Te$_2$ measured at various photon energies from 18.5eV to 23.5eV. The dash lines are theoretical band calculation of IrTe$_2$ (after Fermi energy shifting) at corresponding k$_z$ momentum. The top positions of the valence band (marked by the red dash line) at different k$_z$ confirm the type-II property of Dirac cone in Ir$_{1-x}$Pt$_x$Te$_2$.

**Figure 5| Superconductivity of Ir$_{1-x}$Pt$_x$Te$_2$ under low temperature. a,** Resistance versus temperature of three Ir$_{1-x}$Pt$_x$Te$_2$ samples from x=0.1-0.3 under low temperature. **b,** Temperature dependence of upper critical field of these three samples. The corresponding curves are the best fit lines to the experimental data. **c,** The superconductivity phase diagram of Ir$_{1-x}$Pt$_x$Te$_2$. The brown data point of x=0.04 is extracted from ref. 36. The sketch displays the relative relation between Fermi level (black dash line) and type-II Dirac points in the corresponding Ir$_{1-x}$Pt$_x$Te$_2$ samples measured by ARPES.

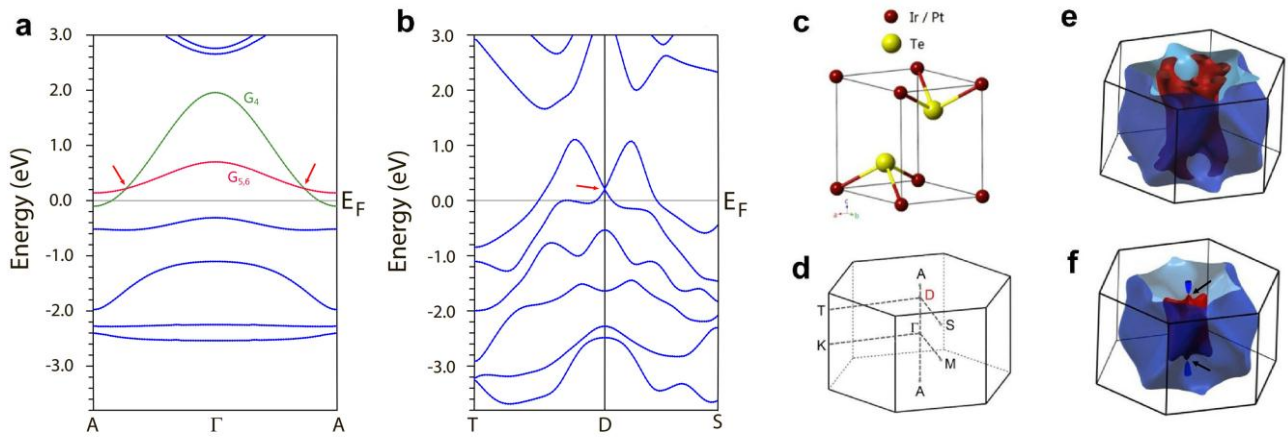

**Figure 1| Ideal type-II Dirac band structures in $Ir_{1-x}Pt_xTe_2$.**

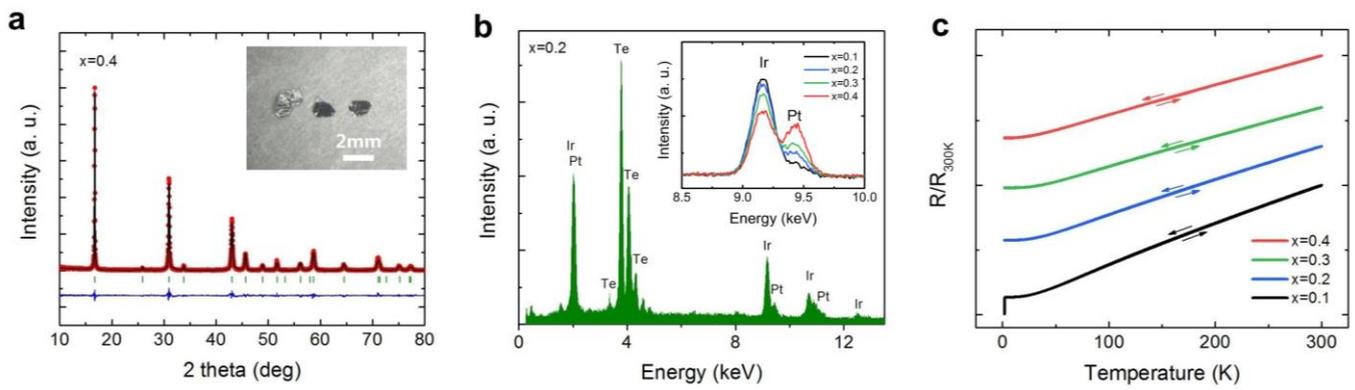

**Figure 2| Crystal growth and characterization.**

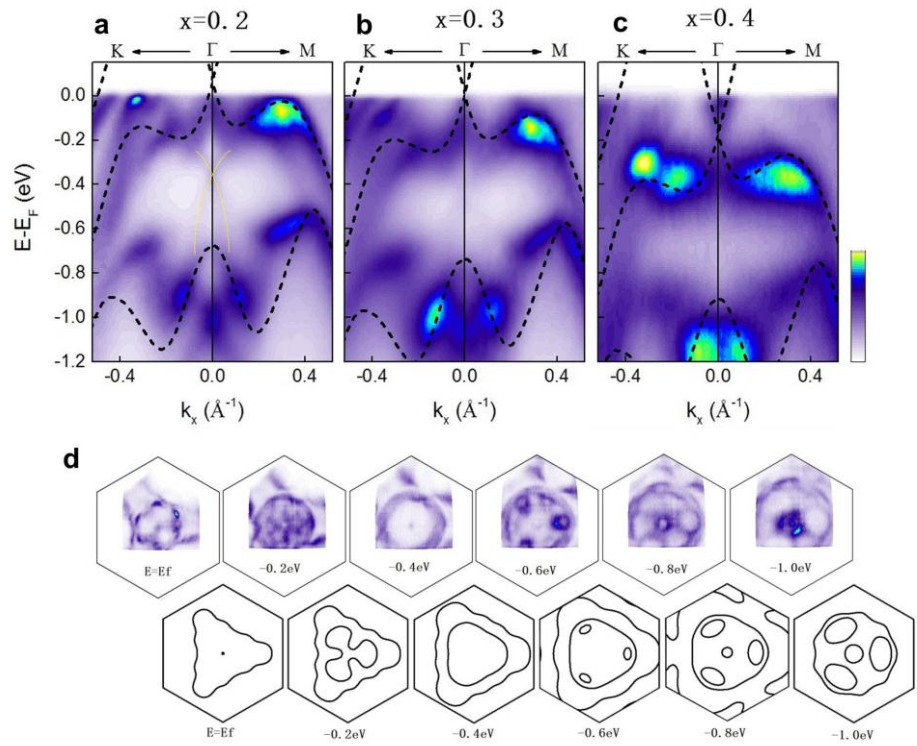

**Figure 3| Dispersions with various doping measured by ARPES.**

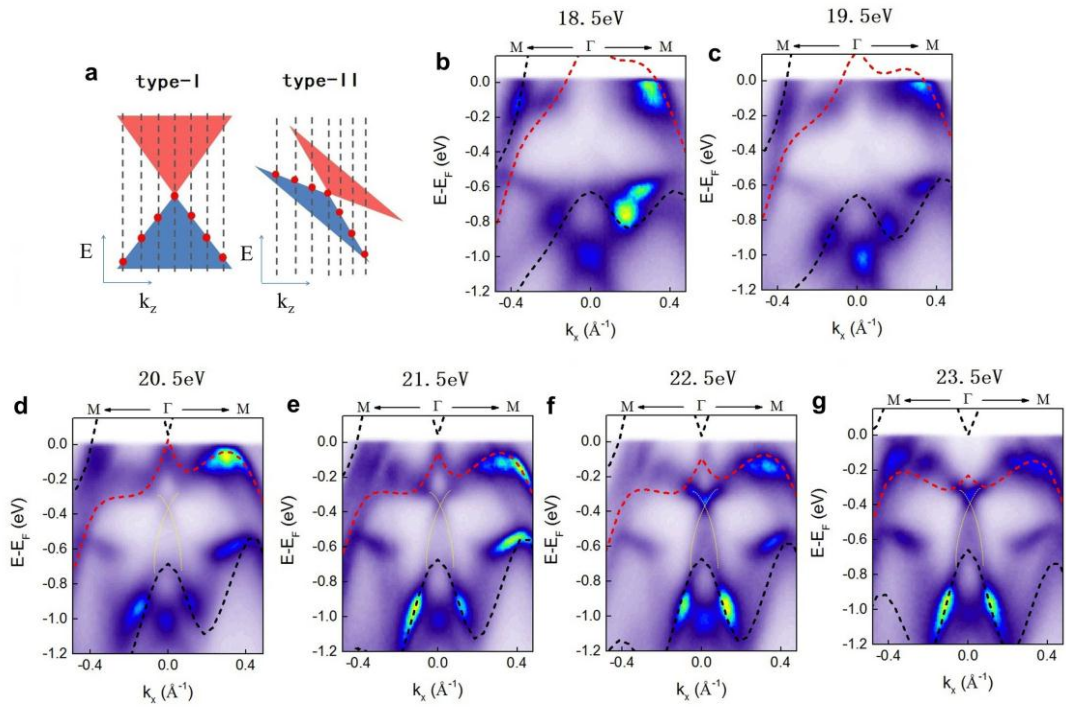

**Figure 4| Verifying type-II Dirac property in Ir$_{1-x}$Pt$_x$Te$_2$.**

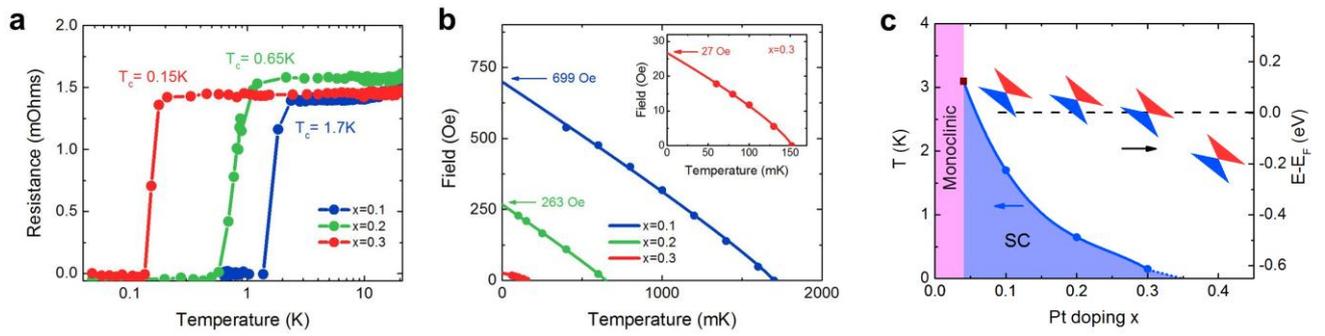

**Figure 5| Superconductivity of Ir$_{1-x}$Pt$_x$Te$_2$ under low temperature.**